# Matching Terahertz and Hall Mobilities as a Hallmark of Intrinsic Charge Transport in Metal-Halide Perovskites

Dmitry R. Maslennikov[1], Ben P. Carwithen[1], Vladimir V. Bruevich[2‡], Yichao Cai[1], Davide Nodari[1], Navendu Mondal[1], Xijia Zheng[1], Beier Hu[1], Nicola Gasparini[1], Jarvist M. Frost[1], Vitaly Podzorov[2†], and Artem A. Bakulin[1*]

*[1]Department of Chemistry and Centre for Processible Electronics, Imperial College London, UK*

*[2]Department of Physics and Astronomy, Rutgers, the State University of New Jersey, New Jersey 08854, USA*

** a.bakulin@imperial.ac.uk; †podzorov@physics.rutgers.edu; ‡bruevich@physics.rutgers.edu*

**Abstract** :

Charge-carrier transport in soft-lattice materials, including metal-halide perovskites, is often perceived to be highly heterogeneous across different length scales, and influenced by both the intrinsic (dynamic) thermal electronic disorder and extrinsic (static) disorder due to crystal defects, impurities, grain boundaries, and surface states. As a consequence, the reported carrier mobilities obtained by different electrical and optical measurement techniques frequently disagree, raising a critical question: can a truly intrinsic charge transport regime (that is, a regime not dominated by static disorder) extend across macroscopic single crystals of these materials? Here, we demonstrate such a regime in an exemplary metal-halide perovskite system, epitaxial $CsPbBr_3$ single crystals, where the local mobility obtained via optical pump–terahertz probe (OPTP) spectroscopy quantitatively agrees with the macroscopic transport mobility across a broad range of experimental conditions. Using a dedicated device platform that enables concurrent Hall-effect and OPTP measurements on the same single-crystalline sample, we obtain consistent room-temperature mobilities of ~ 30 $cm^2V^{-1}s^{-1}$, among the highest reliably reported for $CsPbBr_3$. Both techniques reveal band-like temperature dependence of the hole mobility with similar power exponents, confirming that the same intrinsic transport mechanism governs the ultrafast/local and steady-state/macroscopic responses. These results show that defect-free charge transport is achievable in soft-lattice perovskites on millimetre length scales and establish a robust methodology for benchmarking intrinsic mobility in emerging semiconductors.

## Introduction

The charge carrier mobility, $\mu$, is a key transport parameter directly setting the conductivity and transconductance, and majorly influencing carrier diffusion lengths, and switching speeds of semiconductor devices.[1–5] The accurate determination of mobility remains challenging for "soft" processible semiconductors - materials with mechanically compliant lattices,[6] high dynamic disorder,[7,8] and low-temperature solution or vapour processing routes.[9,10] The mobility reported for the same semiconductor can differ by orders of magnitude, depending on sample morphology, device architecture, and measurement method.[11–15]

Two fundamental sources drive the discrepancies in the reported mobilities. First, charge transport in soft semiconductors is intrinsically heterogeneous across different spatial length scales[16], and within the density of states.[17,18] By its original definition, the charge carrier mobility,  is a coefficient of proportionality between the steady-state (final) drift velocity of a carrier established in an external electric field as the result of averaging a very large number of scattering events taking place over macroscopic length and time scales. Naturally, such "transport" mobility is only meaningful as a macroscale material's property, probed non-locally in steady-state measurements. Experimentally, the transport mobility may exhibit an apparent dependence on the device size due to structural disorder or chemical inhomogeneities. Consequently, the apparent transport mobility may depend on the scales being probed,[16,19] and where within the density of states the carriers are being driven. Second, different measurement techniques interrogate different aspects of the transport process: some are contact-based while others are contactless;[5,20–22] some probe ultrafast local response (as in time-resolved optical techniques), while others capture steady-state macroscopic conduction;[20,22] and each may be preferentially sensitive to surface, interface, or bulk properties.[23–25] In particular, the mobility extracted in ultrafast optical experiments (i.e., "optical" mobility), represents a technique-specific methodological parameter, rather than the actual transport mobility, and can thus lead to additional discrepancies or even confusion when compared to steady-state, macroscale transport measurements.

The intrinsic carrier mobility is the fundamental transport property of a semiconductor determined solely by the carrier effective mass (band structure), dynamic (thermal) disorder, and electron–phonon interactions, without the contribution from extrinsic (static) disorder, such as crystal defects, impurities, interfaces, or morphological barriers.[26,27] This intrinsic mobility represents the maximum charge-transport performance that a material class can, in principle, deliver. In practice, however, real devices almost never reach this limit as the effects of static disorder inevitably reduce the apparent measured mobility below the intrinsic value.[25] Achieving intrinsic mobility in a macroscopic functioning device is therefore highly significant, as it would indicate that the material's full transport potential can be realised on operational length scales — a regime that is rarely observed in soft, processible semiconductors.

Experimentally, the intrinsic limits of the mobility are often inferred from high-quality single-crystal field-effect transistors (FETs),[5,26,28,29] which (when operating within well-defined transport regimes) provide a benchmark for reliable mobility evaluation. Yet even such devices remain sensitive to surface disorder, contact effects, and long-range inhomogeneities, meaning that extracted $\mu$ values typically still fall short of intrinsic mobility. Additionally, anomalies and artifacts in FET characteristics may result in mobility overestimation.[13] Thus, while FETs are regarded as a convenient experimental tool, they can rarely identify the intrinsic mobility limits in emerging materials, unless high-quality single-crystalline FETs are fabricated and carefully measured, following the established protocols of reliable mobility extraction.[12,13,30]

An efficient tool for overcoming some of the limitations of FET measurements is the Hall effect method known to primarily probe band-like delocalized charge carriers, making it helpful when the intrinsic mobility is investigated.[27] In addition, it allows the determination of the type (positive or negative) of the majority carriers.[10,31] Since the first clear demonstration of a Hall effect in soft semiconductors,[32] it has become an important experimental tool for unravelling the intrinsic charge transport properties of various small-molecule organic semiconductor,[33–38] as well as conjugated polymers,[39–44] nanotubes,[45] and perovskites.[10,46] Despite the clear advantages of the Hall technique, it still requires a high-quality FET device, making the application of this technique relatively scarce. Furthermore, there have been reports of misinterpretation of the Hall experiments,[12] apparently stemming from neglecting the save practices of Hall measurements.[30] These factors are responsible for the large spread of the reported $\mu$ values extracted via FET or Hall techniques for the same material.[12]

An alternative route to accessing intrinsic mobility is to use contactless probes that measure transport on nanometre-length scales, where crystallinity variations, interfaces, and extended defects have minimal influence.[47,48] Among such techniques, optical pump–terahertz probe (OPTP) spectroscopy has emerged as a powerful method for quantifying photoconductivity and local (~5-nm)[16] charge-carrier mobility.[4,25,49–62] By interrogating carriers immediately after photoexcitation—before substantial trapping, interfacial scattering, or long-range transport limitations develop—OPTP offers a way to circumvent many of the constraints inherent to device-based measurements.[51] Over recent years, this technique has been widely applied to soft-lattice semiconductors, including organic single crystals, conjugated polymers, chalcohalides, and metal–halide perovskites. Although the distribution of reported mobilities in metal–halide perovskites remains noticeable (typically a few fold rather than orders of magnitude), OPTP measurements are typically more consistent than electrical measurements, because fewer processes perturb the carrier dynamics at the nanoscale, with contactless interrogation allowing to further reduce variations and noise.[11]

In contrast to FET or Hall techniques[63], OPTP spectroscopy is intrinsically time-resolved, enabling direct observation of transient transport phenomena such as carrier cooling,[49,64] localization dynamics,[16] trapping, and polaron formation.[49,65–67] OPTP yields the total mobility of all photocarriers rather than distinguishing between electrons and holes,[51,68] making it a complementary probe to electrical methods that isolate specific carrier types. Nevertheless, several factors limit its ability to reliably extract the mobility. Many OPTP studies are performed on optically thin, poorly crystalline thin films with significant disorder and high defect densities - materials unsuitable for robust electrical characterization, where the link between local and macroscopic transport would be obscured. In addition, variations in thin-film morphology and substrate interactions introduce further uncertainty, hindering reproducibility and complicating direct comparison with device-based mobility measurements.

Therefore, a reliable determination of the carrier mobility in soft-lattice materials must disentangle intrinsic transport from extrinsic effects associated with short- and long-range static disorder and inhomogeneities.[28,69,70] This requires a direct comparison between the macroscopic steady-state (transport) mobility extracted from electrical measurements and the local (optical) mobility obtained by contactless ultra-fast spectroscopic probes, performed on the very same sample, ensuring that both measurements are carried out using the identical material quality, interfaces, device geometry, and boundary conditions. Only when these two mobilities are benchmarked side-by-side, under device-relevant conditions, can one confidently assess how close the material operates to its intrinsic limit. Consequently, there is a critical need for experimental platforms that enable such co-localized electrical and optical mobility measurements and provide a robust basis for accurate mobility determination in the genuinely intrinsic transport regime.

To address these challenges, we developed an integrated platform that enables magneto-transport and OPTP spectroscopic measurements to be performed on the same device. This makes it possible, for the first time in soft-lattice metal-halide perovskites, to establish a direct, quantitative link between the local, ultrafast carrier dynamics and macroscopic steady-state charge transport - something that is rarely achievable when measurements are carried out on different samples.

As a test bed for our studies, we use high-quality epitaxial $CsPbBr_3$ single-crystal devices,[10] a material system known for low trap densities and favourable structural coherence. By extracting the mobilities from both Hall and OPTP measurements in the same sample under identical conditions, we can evaluate not only the accuracy of each method but also whether the material can sustain charge transport close to its intrinsic limit. In our samples, the two mobilities, and their temperature dependences, converge within a narrow range, revealing that single-crystalline $CsPbBr_3$ can operate in a regime where local and macroscopic transport properties coincide. This demonstrates that intrinsic charge transport can be achieved in a soft, processible perovskite and highlights the value of our co-localised characterisation approach for identifying such regimes with confidence.

## Results and discussion

Figures 1 a-b show a photograph of the perovskite device used in this comparative study together with the schematic device structure and measurements principles. The device comprises a macroscopically large individual single crystalline grain of epitaxial $CsPbBr_3$ film grown on a ~ 80 μm-thick mica substrate (Section 4 of the Supporting Information), graphite contacts painted on the perovskite surface (the big current-carrying contacts and additional little electrodes for the Hall voltage measurements), 20 μm-thick gold wires connecting the contacts to a sample holder, and finally a transparent parylene-*N* capping layer[71] conformally encapsulating the entire device.

The sample holder has an aperture at the centre (under the device) that allows measurement of optical transmission, provided that a sufficient amount of light can pass through the $CsPbBr_3$ crystal, mica, and parylene-*N* layers. The impact of the parylene-*N* capping layer on the THz OPTP signal has been investigated and found to be insignificant (Section 5 of Supporting Information), thus allowing us to conclude that its effect on the intrinsic charge carrier mobility evaluation with the OPTP method is negligible.

Upon completion of the epitaxial growth, the mica substrates are nearly completely covered with crystalline $CsPbBr_3$ domains (Section 6 of the Supporting Information, [10]), each several millimetres across, indicating the high quality of the epitaxial growth. Devices are then prepared on an individual single-crystal grain by mechanically removing all but the selected grain from the substrate with a razor. Importantly, there are some regions of a bare mica substrate, lacking the perovskite film, that occur in these samples (Figure 1a); these regions can be used as a reference for quantifying the background THz absorption of the device.

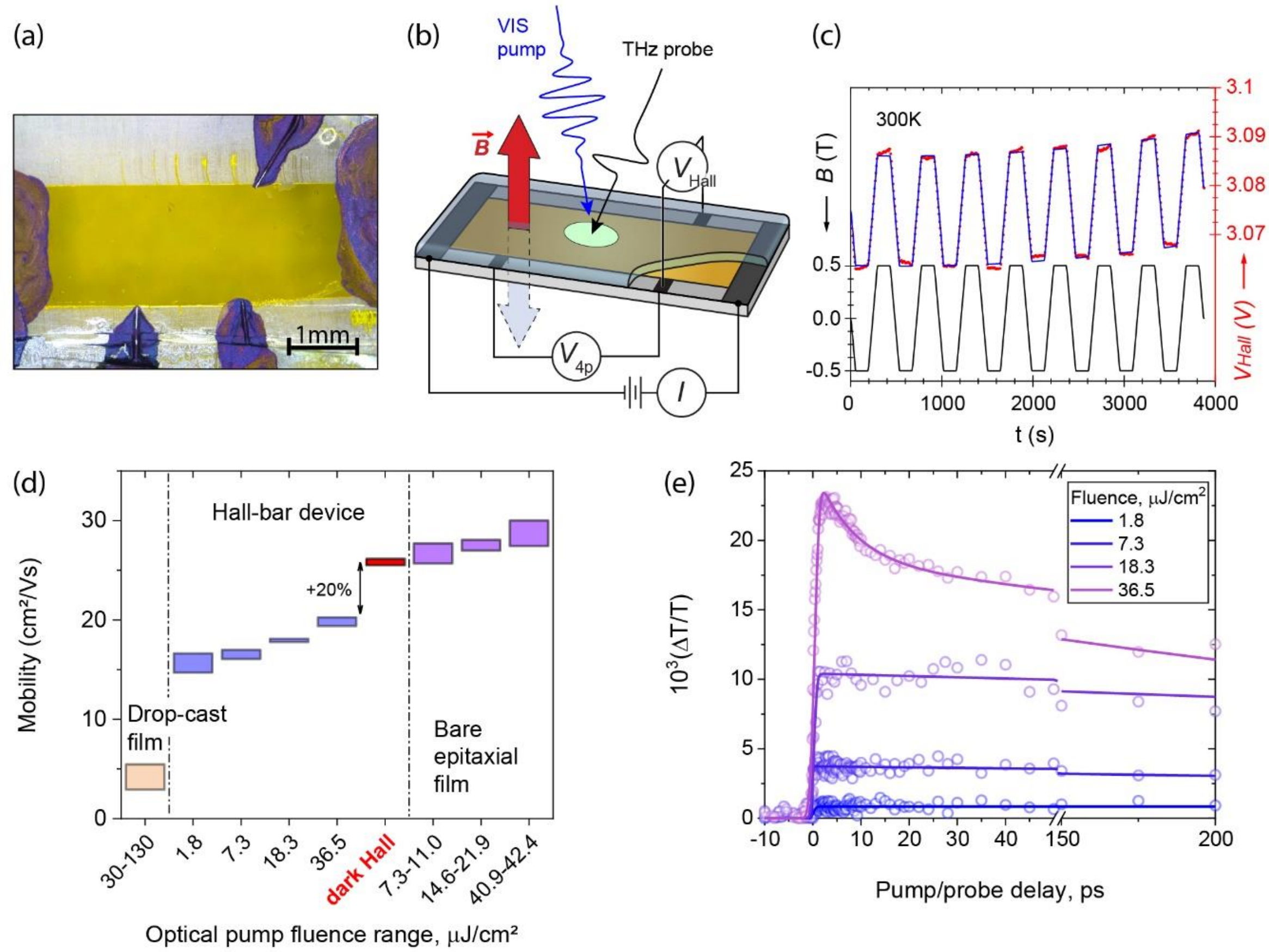

**Figure 1.** (a) Optical microscopy image of the studied $CsPbBr_3$ device with graphite contacts and parylene-N capping layer that allows both the (dark) Hall effect and THz OPTP measurements of the same device. (b) A sketch of the device structure and an overview of the Hall effect and OPTP measurements. The approach allows both electrical transport and transmission spectroscopy measurements in a single device. (c) An example of the Hall voltage measurements of the device shown in panels a and b carried out at 300 K in the dark, revealing an excellent signal-to-noise ratio: the transport (Hall) mobility is extracted from the modulations of the Hall voltage induced by the external magnetic field, while the longitudinal current ($I$ = 15.3 μA) is flowing through the sample, and the longitudinal voltage drop ($V_{4p}$ = 7.53V) is simultaneously measured away from the current injection terminals. (d) The mobilities determined with OPTP and Hall methods in three different types of $CsPbBr_3$ samples. $\mu_{OPTP}$ was measured at several fluences to address possible trap-filling effects by photocarriers. (e) Examples of OPTP transients of the single-crystal device shown in panels a and b recorded at several fluences.

This study is based on simultaneous application of steady-state electrical transport and fully optical ultrafast characterization techniques performed on the same, large-area, single-crystalline perovskite device featuring a mildly conducting perovskite-parylene-*N* interface.

In the optical part of the experiment (OPTP measurements), the two ultrafast optical pulses are used to generate carriers and probe their dynamics. First, charges are photoexcited in the active area of the device by an ultrafast optical pump pulse. The photoexcitation is followed by a terahertz pulse, transmission of which is sensitive to the density and mobility free charges photoexcited in the semiconductor.[51] THz was detected via electro-optic sampling and the maximum of the THz field was used to set up the probe-detection pulse timing. The pump pulse is passed through an optical chopper, allowing the measurement of the difference in terahertz pulse transmission with and without photoexcitation of charge carriers (with and without pump pulse). These changes in transmission are monitored as a function of time delay between optical pump and THz probe pulses, which is proportional to free carrier concentration in time. This means that charge carrier dynamics can be effectively extracted by monitoring the relative change in the THz pulse transmission (Figure 1e). $\Delta T/T$ transients are then used in combination with the charge-carrier density at a particular pump-probe delay to calculate the charge carrier mobility $\mu_{OPTP}$ [4,25,50–52,59] (Section 6 of the Supporting Information).

OPTP transients for the studied device are shown in Figure 1d. The actual $\Delta T/T$ values used for mobility calculations were taken at the time when free mobile carrier formation is completed[72], but bimolecular (electron-hole) recombination has not started yet (typically 2-4 ps after the pump pulse). The values of the extracted mobilities are shown in Figure 1c. We measured OPTP mobilities of several types of $CsPbBr_3$ samples, including the neat epitaxially-grown single-crystal films on mica and more disordered drop-cast films deposited onto THz transparent z-cut quartz substrates.

Figure 1d demonstrates a similar trend for all the samples measured with OPTP: an increase in the extracted mobility with the excitation density (fluence). Upon the increase in fluence, the mobility saturates, which is indicative of trap filling or interfacial doping.[73–75] In the experiments relying on carrier photoexcitation (including the THz OPTP), an internal photo-carrier excitation (IPCE) quantum efficiency, $0 \leq \phi \leq 1$, for free carrier generation must be introduced to connect the apparent measured optical mobility $\mu_{OPTP}$ with the intrinsic microscopic mobility $\mu$: $\mu_{OPTP} = \phi\mu$. Therefore, OPTP provides a lower-bound estimate of the intrinsic mobility. As the photoexcitation fluence increases, $\phi$ approaches unity, consistent with reduced influence of trapping processes at higher carrier densities. Hence, $\mu_{OPTP}$ obtained at higher fluences is more representative of the intrinsic transport mobility extracted from steady-state transport measurements of single-crystalline devices. This effect is consistent with our magneto-transport measurements of these devices. In particular, we have systematically observed that at very long durations typical for steady-state photo-Hall effect measurements, electrons in $CsPbBr_3$ crystals appear to be fully trapped,[10] and only holes contribute to the Hall voltage under *cw* photoexcitation. The neat (uncapped and insulating) single-crystalline $CsPbBr_3$ sample exhibited a much higher OPTP mobility $\mu_{OPTP} = 27 - 30$ cm$^2$V$^{-1}$s$^{-1}$) compared to the solution-cast polycrystalline film ($\mu_{OPTP} < 5$ cm$^2$V$^{-1}$s$^{-1}$) and only marginally higher than that obtained in a parylene-N capped device based on the same kind of epitaxial single-crystalline $CsPbBr_3$ samples ($\mu_{OPTP} \approx 20$ cm$^2$V$^{-1}$s$^{-1}$). The latter device was also used in the Hall-effect measurements performed in the dark ($\mu_{Hall} = 25.8 \pm 0.3$ cm$^2$V$^{-1}$s$^{-1}$), possible thanks to a mild hole doping of the parylene-N/$CsPbBr_3$ interface. To our knowledge, these mobilities are some of the highest values among the mobilities reliably measured in metal-halide perovskites.

Exactly the same epitaxial single-crystalline device (as the one shown in Fig. 1a) was used to perform the Hall-effect measurements in the dark: the pair of big contacts (on the right and left of the patterned yellow film) inject/drain charge carriers into/from the active layer, while little contacts on the sides of the channel are used to probe the transverse (that is, Hall) and the longitudinal (four-probe) voltages, thus ensuring contact-corrected (four-probe) measurements of the sample's conductivity and Hall mobility. The charge injection itself is facilitated by a mild charge-transfer doping of $CsPbBr_3$ surface occurring at the interface with parylene-N and leading to a conductivity of ~ 2 μS/square even in the dark. The injected holes drifting along the channel experience a transverse Lorentz force in a magnetic field applied perpendicular to the film, leading to a Hall voltage between the Hall probes. In this case, for a more reliable extraction of the Hall mobility, the magnetic field was slowly swept between -0.5 and 0.5 T, with the Hall mobility extracted from the corresponding modulations of the Hall voltage (Figure 1c). In addition, very slow parasitic drifts and fluctuations in the recorded Hall voltage, typical of low-level Hall measurements in resistive materials, were taken into account by performing a polynomial fit and background subtraction (Section 6 of Supporting Information). The Hall mobility measurements in this device in the dark yield $\mu_{Hall} = 25.8$ cm$^2$V$^{-1}$s$^{-1}$ (Figure 1d).

Carrying out the OPTP measurements on the same device resulted in $\mu_{OPTP} = 19.8 \pm 0.4\ \mathrm{cm^2V^{-1}s^{-1}}$ (at $36.5\ \mu J/cm^2$). Although there is a minor mismatch between the observed $\mu_{Hall}$ and $\mu_{OPTP}$, the consistency between the results obtained by these two dissimilar techniques, probing the system at different time and length scales, is remarkable. To our knowledge, this is the closest agreement between carrier mobilities measured by an ultrafast optical/spectroscopic probe and a steady-state macroscopic charge transport technique in a soft-lattice material. The minor mismatch (of about 20-30%) may originate from several sources. First, OPTP yields a weighted total of the electron and hole mobilities, $\mu_e$ and $\mu_h$, which can be comparable but not necessarily equal to each other,[76] while the (dark) Hall effect in these devices only probes holes induced at the perovskite/parylene-N interface. Second, OPTP is predominantly a bulk measurement, while the (dark) Hall measurements probe interfacial holes that "see" the low-dielectric-constant environment (parylene-*N*) that could be affecting their hole mobility via the surface Froehlich polaron mechanism.[77] Third, tail states near the mobility edge, formed due to the strong dynamic disorder in soft-lattice perovskites, may lead to a noticeable population of hopping carriers coexisting with band carriers.[78] In systems with significant off-diagonal thermal disorder, such a mixed transport regime can exhibit a Hall mobility exceeding the longitudinal transport mobility. Finally, statistical averaging of the carrier relaxation time can lead to a Hall carrier density that somewhat underestimates the actual carrier density (and, conversely, a Hall mobility that slightly overestimates the transport mobility), with the specific correction factor (or, the so-called Hall scattering factor) that depends on the band structure and dominant scattering mechanisms.[79,80] Understanding the origin of the minor difference between $\mu_{OPTP}$ and $\mu_{\text{Hall}}$ observed here in epitaxial crystalline $CsPbBr_3$ requires further investigation.

To gain further insights into the charge transport mechanism, we measured the temperature dependence of the mobilities obtained with OPTP and Hall-effect techniques. Figure 2a shows the comparison of $\mu(T)$ dependencies obtained with obtained with these two dissimilar methods. We found very similar trends in the temperature dependence of $\mu_{OPTP}$ and $\mu_{Hall}$, which can be best described by the power law, $\mu(T) \propto T^{-\text{b}}$. Despite being device-based, the Hall-effect measurements were found to systematically yield slightly higher mobilities with the power exponent of $b = 1.29 \pm 0.02$ (compared to $b = 1.10 \pm 0.03$ in OPTP measurements). Because Hall-effect measurements are known to be very sensitive to various types of defects and disorder, this observation suggests a superb structural quality of the sample over its macroscopic length scale. The observed minor difference in the power exponent can be attributed to the technical procedure of extracting the Hall mobility (Section 6 of Supporting Information) or difficulties with precise determination of the pump's spot size inside of the cryostat in the OPTP experiment. We also note that it is typically more challenging to perform variable-temperature measurements with devices relying on carrier injection from contacts because of the contact issues that may arise at low temperatures (< 200 K). Because of this, the $\mu_{OPTP}(T)$ was recorded in a wider temperature range, between 125 and 300 K.

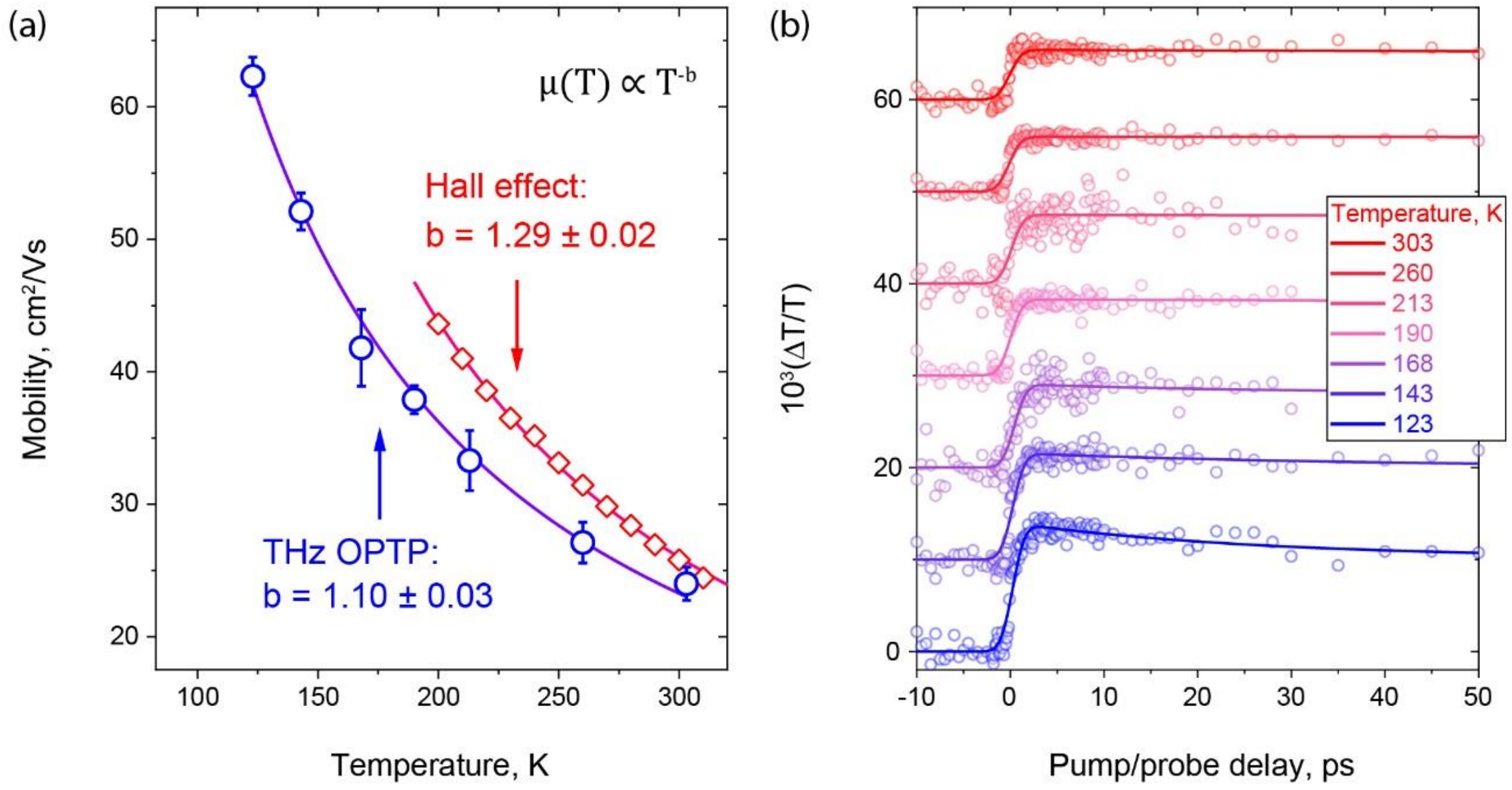

**Figure 2.** Temperature-dependent measurements of the carrier mobility in epitaxial single-crystal $CsPbBr_3$ perovskite films. (a) The mobility extracted from OPTP and Hall measurements in the ranges 125 – 300 K and 200 – 315 K, respectively. The $\mu(T)$ follows a power-law dependence, $\mu(T) \propto T^{-\mathrm{b}}$, with the power exponent indicated on the plot (Pearson's correlation coefficient $R$ = 0.99 in both cases). (b) The OPTP transients corresponding to the extracted $\mu_{OPTP}(T)$ recorded at a fluence of ≈ 8.4 μW/cm² (excitation pulse power ≈ 100 μW) and subsequently fitted by a sum of two exponential functions convolved with a Gaussian.

The observed behaviour ($\mu(T) \propto T^{-\mathrm{b}}$, b > 0) is indicative of a band-like transport, in which mobility is limited by phonon scattering. Recently, theoretical modelling has been carried out investigating the exact value of the temperature power exponent. For large polarons with the mobility limited by longitudinal optical (LO) phonon emission, $\mu(T)$ dependence is expected to approximately follow a $T^{-0.5}$ dependence,[81] whereas acoustic-phonon scattering and other thermally activated phonon scattering processes in a semiclassical treatment lead to a $T^{-1.5}$ dependence.[82] Here we obtained the result that is closer to the classical dependence that has been predicted earlier for Drude-like delocalized charge carriers. This fact is corroborated by the photoconductivity measurements showing a typical Drude-like response (see Section 7 of the Supporting Information). The temperature dependence of the mobility is only circumstantial evidence of a dominant scattering process[83], and careful consideration of different phonon energies and scattering matrix elements are required for precise knowledge of the factors limiting intrinsic mobility. Recent modelling[84] with the Boltzmann transport equation (i.e., considering semi-classical discrete scattering events) reproduces the observed $\sim T^{-1.5}$ temperature dependence fairly well, thus suggesting that a Drude-like theory is sufficient to model the intrinsic charge transport measurements in crystalline $CsPbBr_3$.

Figure 2b shows the OPTP kinetics (the $\Delta T/T(t)$ transients) used for the mobility extraction. At higher temperatures (> 140 K), we observed almost no changes in $\Delta T/T(t)$ ($\sim n_{charges}$) over the measured time interval (300 ps). Therefore, for those temperatures, the $\Delta T/T(t)$ transients can be approximated by a single monomolecular decay function (one exponent in the convolution, see SI for fitting details). At lower temperatures (< 150 K), a faster decay is clearly observed within the first ~ 20 ps, suggesting that an additional faster exponential decay component ($\tau_2 \sim$ 10 – 30 ps) is necessary to fit the data (see SI for fitting details). An increase in the rate of charge recombination with lowering temperature is generally expected in semiconductors. As this work is concerned with accurate evaluation of the carrier mobility, it would be imperative to understand if the changes observed in the kinetics influence the extraction of $\mu$. Usually, slight changes in recombination kinetics are not of concern for mobility extraction, if the recombination times are within the temporal resolution of the setup. Here, for example, we used $\Delta T/T$ averaged in the 2 – 4 ps time window, corresponding to the maximum of the THz modulation in the OPTP kinetics, where we expect the photon-to-charge yield to be close to unity.

However, at higher excitation fluences, when much faster recombination rates can influence OPTP transients, the assumption of 100% IPCE can be inaccurate, leading to inaccuracies in the extracted $\mu_{OPTP}$. To evaluate excitation fluence ranges that can be safely used in OPTP measurements for an accurate mobility extraction, we have performed fluence dependent measurements at both high (300 K) and low (93 K) temperatures.

Figure 3a and 3b shows the fluence-dependent OPTP kinetics recorded at 300 K and 93 K, respectively, together with their approximation curves. It can be seen that the $\Delta T/T$ kinetics at high excitation fluences are dominated by a faster decay component. Figure 3c shows the $\Delta T/T$ values at 300 ps as a function of excitation fluence for both temperatures. Looking at the 300 K curve, it can be clearly seen that the $\Delta T/T$ signal scales linearly with the fluence; however, the slope of the linear dependence changes at $f_1 = 40\ \mu J/cm^2$, which suggests a change in the charge recombination regime. A similar trend can be seen in the 93 K data, although the transition point for this temperature is shifted to lower fluences, with the kink occurring at $f_2 = 9\ \mu J/cm^2$. These fluences coincides with the appearance of the faster components in Figure 3a and 3b, supporting the conclusion of a change in the charge carrier recombination regime. We have also extracted $f_1$ and $f_2$ values from the fits, which coincided very well with the values presented here (see Section 9 of the Supporting Information).

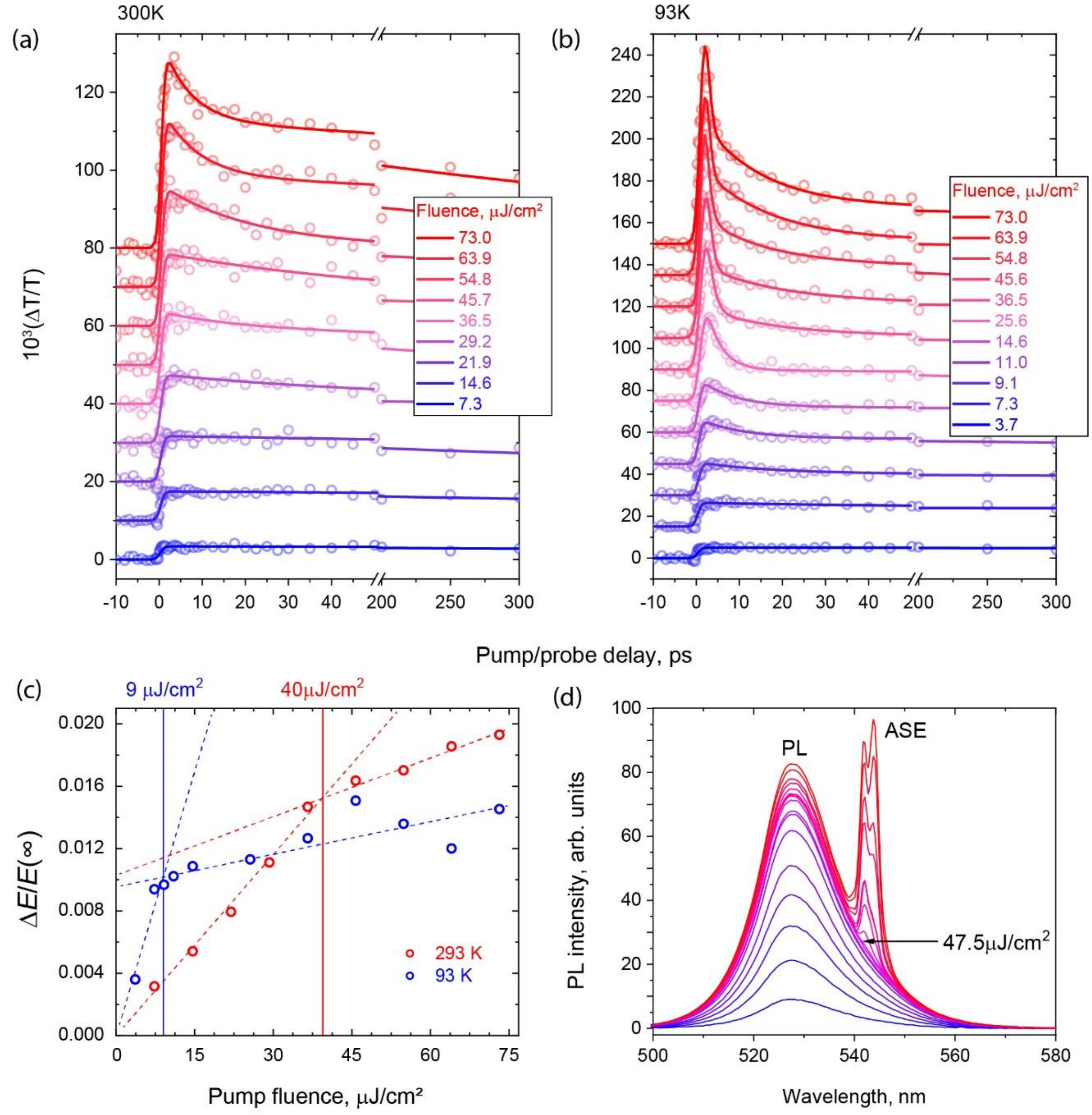

**Figure 3.** (a, b) Fluence dependence of the OPTP transients in epitaxial single-crystalline $CsPbBr_3$ films under a 400 nm excitation recorded at 300 and 93 K. (c) The OPTP signal at 300 ps after photoexcitation as a function of excitation fluence for these temperatures. The kink observed on the curves at approximately 9 and 40 μJ/cm$^2$ signal a change in the charge carrier dynamics and recombination regime. (d) Fluence dependant PL spectra of the same sample under the same photoexcitation (400 nm, 35 fs laser pulses).

We found that the probable cause of the change in the THz dynamics at high excitation fluxes is the emergence of amplified stimulated emission (ASE) that reduces the charge carrier concentration. Figure 3d shows the PL of an epitaxial single-crystalline $CsPbBr_3$ film as a function of the laser excitation fluence recorded at 300 K. At low fluences, the PL spectra are centered at around 530 nm. Upon increase in the fluence, a sharp narrow peak appears at 540 nm, which very well coincides with the emergence of ASE reported for $CsPbBr_3$ in the literature.[85,86] Moreover, the observed ASE threshold in the static PL, $f_{PL}^{ASE} \approx 47.5\,\mu J/cm^2$(Figure 3d), coincides well with the kink observed in THz OPTP measurements at $f_1 = 40\,\mu J/cm^2$. Similarly, the kink point for the THz measurements at 93 K ($f_2 = 9\,\mu J/cm^2$) closely matches the ASE threshold determined in low-temperature PL measurements (see Section 3 of the Supporting Information).

The onset of the stimulated emission has a simple explanation in terms of the Mott physics of polaron states: it is at this density that the hole and electron polarons are forced into overlap. Therefore, the polarisation shell which protected the polarons from recombination is lost, and the Einstein A and B coefficients discretely jump up. The charge carriers in this state are no longer well-defined polaronic quasi-particles with a renormalised effective mass, but rather an electron-hole polaron plasma. The concept of mobility as a fundamental parameter defined in the linear response theory, is no longer relevant when one induces collective excitations.

From the 300 K turn-on of ASE at 47.5 μJ/cm$^{-2}$, we calculate the density of charge carriers to be 6.6x10$^{18}$ cm$^{-3}$, and from the inflection in THz transmissivity at 40 μJ/(cm$^{-2}$) we calculate the density as 5.8x10$^{18}$ cm$^{-3}$. These figures agree well with ~10$^{18}$ cm$^{-3}$ ASE threshold calculated with the Feynman variational approach and a Frohlich Hamiltonian for $CsPbBr_3$ (see Section 6 of the Supporting Information). As described in Ref [87], the polaron gets larger at lower temperatures, and we estimated this increase to be from 42 Å at 300 K to 56 Å at 93 K, which leads to a predicted 93 K threshold density to be 0.4x10$^{18}$ cm$^{-3}$ (corresponding to the pump density of 2.6 μJ/cm$^2$), slightly lower than the observed threshold. These estimates show that the simple polaron physics of the material can be used to interpret the observed crossovers (kinks) and provide quantitative predictions in agreement with experiment by solving a model Hamiltonian parameterized with the bulk material properties.

One intriguing possibility for future studies is that these measurements may offer an insight into the electron-hole polaron plasma. Theoretical understanding of the behaviour of such transient population of charge carriers is currently lacking, as it is a many-body finite-temperature quantum mechanical system, where many of our theoretical methods (such as the single-polaron low-density limit used in the Feynman variational theory) are not applicable.

We therefore identified the acceptable maximum fluences for reliable carrier mobility measurements with OPTP-based techniques as $f_1 = 40\,\mu J/cm^2$ at 300 K, and $f_2 = 9\,\mu J/cm^2$ at 93 K. Overall, we found that the $\Delta T/T$ at the transient curve's maximum (at 2-4 ps) does not scale proportionally with the fluence after reaching ASE level, which leads to underestimation of the charge carrier mobilities (see Section 8 of the Supporting Information). We anticipate that several factors might be involved in this result. First, fast charge carrier recombination may exceed the temporal resolution of the THz OPTP setup, taking into account the instrument response function of ~1 ps due to the THz pulse duration. This limitation is very likely to occur under high-fluence conditions, where ultrafast processes have been reported in perovskites, including superluminescence with extremely rapid decay rates.[88–90] Second, at high excitation densities, carrier mobility may also be reduced by enhanced many-body interactions, including carrier–carrier scattering.[91] Such density-dependent scattering would further contribute to the apparent decrease in mobility extracted above the ASE threshold.

In order to understand practical factors affecting the charge carrier mobility in OPTP measurements of $CsPbBr_3$, we have performed a range of OPTP measurements using separate samples (or batches of samples) of various kinds, including the neat (as-grown) and aged epitaxial single-crystal films, as well as disordered films drop-cast from solution. In particular, we studied how the mobility is affected by (a) moving from grain to grain on the same large epitaxial $CsPbBr_3$/mica sample (checking spatial isotropy), (b) age of these epitaxial films, and (c) the level of disorder and film morphology by repeating measurements on solution-cast samples. In respect to (a), we found a great spatial uniformity when different domains are measured at distant points of the same large sample, with the variations in $\Delta T/T$ of less than 10% (Section 1 of the Supporting Information). In respect to (b), when testing the effects of sample ageing, we measured several epitaxial samples grown within a few months of each other and found that they yielded similar results. This indicates that highly crystalline $CsPbBr_3$ films retain good stability over many months, especially considering they were unencapsulated (in these tests) and stored in ambient laboratory air. In a control measurement of a sample that was more than two years old at the time of the test, we observed a reduction in mobility by roughly a factor of three (7.7 ± 0.5 $cm^2V^{-1}s^{-1}$) (Section 2 in the Supporting Information). However, the lower mobility in this case may also be attributed to other factors, such as slightly different growth conditions used in earlier batches. To address (c), the above results for $\mu_{OPTP}$ of highly crystalline films were also compared with the films made by drop casting from a solution (Figure 1c, Fig. S2). The drop-cast films have been found to have optical mobilities several times lower (≈ 4 $cm^2V^{-1}s^{-1}$), and we associate this with the challenges normally associated with this type of sample preparation, such as lower crystallinity and defects incorporation. Further post-processing, such as solvent vapour annealing may help to improve the overall quality of the drop-cast films.

Simultaneous ultrafast optical and steady-state magneto-transport measurements are beneficial, as they allow to test the reliability of each method, as well as verify the accuracy of the mobility estimation for the particular material. In this case, we found our values for $\mu_{OPTP}$ and $\mu_{Hall}$ to be among the highest values reported for $CsPbBr_3$-based perovskite devices. Usually though, this is not the case, and mobilities identified with spectroscopic methods are typically much higher than those reported in steady-state macroscopic transport measurements of devices (even though methodological errors can lead to unrealistically high mobility estimates in both optical and electrical measurements).[11,13] It is generally accepted that differences between $\mu_{Hall}$ and $\mu_{OPTP}$ are associated with the local and transient character of ultra-fast spectroscopic probes.[12,16] The $\mu_{OPTP}$ is less sensitive to structural and morphological crystal defects, because of the local and ultrafast character of these measurements. The $\mu_{OPTP}$ is estimated at a few hundreds of femtoseconds after the photoexcitation, and the charges have no time to reach trap sites, grain boundaries, or other defects that would otherwise be a limiting factor for the average mobility. Charge trapping is reported to affect charge-carrier dynamics on a nanosecond timescale. On the other hand, $\mu_{Hall}$ is the result of device-based, steady-state transport measurements over macroscopic length scales (the contact-to-contact distance in our $CsPbBr_3$ Hall-bar devices is approximately 4 mm, Figure 1a), which makes this methodology detrimentally sensitive to defects (especially, fine cracks or grain boundaries), since it relies on a slow drift of carriers over macroscopic distances between the contacts. Therefore, the observed match of mobilities, $\mu_{Hall} \approx \mu_{OPTP}$, suggests that the transport mobility in the epitaxial singe-crystalline $CsPbBr_3$ devices is not limited by defects at any length scale, thus highlighting high quality of the epitaxial perovskite single-crystalline films.

## Conclusion

This work introduced an integrated device platform that enabled truly co-localized optical pump–terahertz probe (OPTP) spectroscopy and steady-state Hall-effect measurements to be carried out on the same high-quality epitaxial $CsPbBr_3$ single-crystal devices. By probing local and macroscopic transport under identical conditions, we obtained mutually consistent room-temperature mobilities ($\mu_{OPTP} = 19.8 \pm 0.4$ $cm^2V^{-1}s^{-1}$; $\mu_{Hall} = 25.8 \pm 0.3$ $cm^2V^{-1}s^{-1}$), which are among the highest reliably reported values for this material. Their close agreement shows that charge transport across the entire macroscopic device is not limited by grain boundaries, interfaces, or other extended defects, and that $CsPbBr_3$ operates in a transport regime close to its intrinsic limit. Temperature dependences ($\mu \propto T^{-b}$, $b \approx 1.1–1.3$) and Drude-like photoconductivity further confirmed band-like, phonon-limited transport.

Beyond demonstrating intrinsic-like behaviour, this platform allowed us to establish practical guidelines for accurate OPTP mobility extraction. Measurements remained reliable below ASE thresholds (≤ 40 μJ $cm^{-2}$ at 300 K; ≤ 9 μJ $cm^{-2}$ at 93 K), before the change of slope in $\Delta T/T$ signal. Higher fluences led to mobility underestimation due to ultrafast recombination and enhanced scattering. Additional tests verified excellent spatial uniformity (<10% variation), long-term stability, and processing-dependent differences, with drop-cast films exhibiting substantially lower mobilities ≈ 4 $cm^2$ $V^{-1}s^{-1}$.

This work establishes a rigorous framework for benchmarking contact-free OPTP spectroscopy against device-based transport metrics. When calibrated on a shared device platform, OPTP provides a reliable proxy for intrinsic mobility and offers a broadly applicable strategy for reproducible transport assessment in soft-lattice, processible semiconductors. The methodology developed here opens a pathway for identifying intrinsic transport regimes in a variety of emerging materials and for advancing mobility metrology across the wider landscape of soft semiconductor technologies.

## Acknowledgements

This project received funding from the European Research Council (ERC) under the European Union's Horizon 2020 research and innovation program (Grant Agreement 639750/VIBCONTROL) and from UKRI/EPSRC (ActionSpec, grant ref: EP/X030822/1). DM acknowledges funding by Imperial College London President's Scholarships. VP and VB acknowledge funding from the U.S. Department of Energy, Office of Basic Energy Sciences, Division of Materials Sciences and Engineering under Award DE-SC0025401 (material synthesis, device fabrication, and measurements), and the Rutgers University's 2024 Research Council award (equipment upgrades).